\documentclass[aip,graphicx]{revtex4-1}
\usepackage{graphicx}
\usepackage{soul}
\usepackage{bm}
\usepackage{amsmath}
\draft % marks overfull lines with a black rule on the right

\begin{document}

%\preprint{}

\title{Data-driven wake model parameter estimation to analyze effects of wake superposition} %Title of paper



\author{M.J. LoCascio}
\email{locascio@stanford.edu}
\affiliation{Civil and Environmental Engineering, Stanford University}
 
\author{C. Gorl\'{e}}%
\affiliation{ 
Civil and Environmental Engineering, Stanford University
}

\author{M.F. Howland}
\affiliation{%
Civil and Environmental Engineering, Massachusetts Institute of Technology
}
% \author{}
%\thanks{}
% \affiliation{}


\date{\today}

\begin{abstract}
Low-fidelity wake models are used for wind farm design and control optimization. 
To generalize to a wind farm model, individually-modeled wakes are commonly superimposed using approximate superposition models.
Wake models parameterize atmospheric and wake turbulence, introducing unknown model parameters that historically are tuned with idealized simulation or experimental data and neglect uncertainty.
We calibrate and estimate the uncertainty of the parameters in a Gaussian wake model using Markov chain Monte Carlo (MCMC) for various wake superposition methods.
Posterior distributions of the uncertain parameters are generated using power production data from large eddy simulations (LES) and a utility-scale wake steering field experiment.
The posteriors for the wake expansion coefficient are sensitive to the choice of superposition method, with relative differences in the means and standard deviations on the order of $100\%$.
This sensitivity illustrates the role of superposition methods in wake modeling error.
We compare these data-driven parameter estimates to estimates derived from a standard turbulence-intensity based model as a baseline.
To assess predictive accuracy, we calibrate the data-driven parameter estimates with a training dataset for yaw-aligned operation.
Using a Monte Carlo approach, we then generate predicted distributions of turbine power production and evaluate against a hold-out test dataset for yaw-misaligned operation.
Compared to the deterministic predictions of the baseline parameter estimates, we find that the MCMC-calibrated parameters reduce the total error of the power predictions by roughly $50\%$.
An additional benefit of the data-driven parameter estimation is the quantification of uncertainty, which enables physically-quantified confidence intervals of wake model predictions.

\end{abstract}


\maketitle %\maketitle must follow title, authors, abstract and \pacs

\section{Introduction}

Wind energy accounted for about 7\% of the global electricity supply in 2021 and is projected to provide up to a third of total electricity supply by 2050 \cite{veers_grand_2019, international_energy_agency_global_2021}.
The performance of wind farms can be substantially impacted by wake interactions between turbines when they are sited within fifteen rotor diameters in the prevailing wind directions \cite{meyers_optimal_2012}, which can result in annual energy losses above 20\% \cite{barthelmie_modelling_2009}.
Numerous studies have documented and analyzed these interactions in the field \cite{sun_review_2020} and in high-fidelity simulations \cite{lange_modelling_2003, porte-agel_numerical_2013, ghate_modeling_2015, wu_modeling_2015}.

Reliable, computationally-efficient predictive modeling of wind farm flow is needed for decision-making for turbine siting, turbine design, and farm operation since computational fluid dynamics are currently prohibitively expensive in these applications \cite{shakoor_wake_2016, meneveau_big_2019}.
Analytical wake models are inexpensive, low-fidelity tools to predict wake interactions for optimization applications such as control systems and plant design \cite{parada_wind_2017, howland_wind_2019, stanley_massive_2019, fleming2019initial, howland_collective_2022}.
Wake models predict the wake-induced velocity deficit relative to the free-stream incident wind for an individual wind turbine \cite{jensen_note_1983, frandsen_analytical_2006, bastankhah_new_2014}.
To reduce computational expense, wake models parameterize atmospheric boundary layer (ABL) and wake turbulence with a wake spreading coefficient \cite{stevens_flow_2017}.

To enable predictions of the velocity deficit trailing a wind turbine, the unknown wake spreading rate must be prescribed.
The traditional approaches to estimate the unknown wake expansion parameter have been to parameterize it as a function of key physical variables \cite{lissaman_energy_1979, emeis_simple_2009, yang_computational_2012, stevens_flow_2017}, such as friction velocity, or to couple the wake model with a top-down wind farm model \cite{stevens_coupled_2015, starke2021area}. 
Another more recent approach has been to assume the wake spreading rate depends only on turbulence intensity and to tune a linear function using comparisons to idealized large eddy simulations (LES) \cite{niayifar_analytical_2016} or utility-scale wind turbine \cite{teng_calibration_2020} data. 
Once the wake spreading rate is prescribed, the predictions from analytical wake models have been compared to wind farm data from LES \cite{porte-agel_numerical_2013, stevens_generalized_2016} and utility-scale farms \cite{carbajo_fuertes_wind_2018, hamilton_comparison_2020,doekemeijer2022comparison}.
 
The predictions of these empirical models are sensitive to errors in the tuned parameters in addition to inherent modeling error.
Recent studies have demonstrated that wake models that were tuned using data from idealized neutral ABL conditions may lead to predictive bias for conditions outside of the tuning dataset \cite{doekemeijer_closed-loop_2020, howland_optimal_2020, schreiber_improving_2020, howland_optimal_2022}.
Further, analytical wake models continue to increase in complexity to include additional physical effects such as yaw-induced wake deflection \cite{bastankhah_experimental_2016, howland_wake_2016}, wake three-dimensionality \cite{bastankhah_new_2014, shapiro_modelling_2018, martinez-tossas_aerodynamics_2019}, shear \cite{gebraad_incorporating_2016}, and veer \cite{abkar_analytical_2018}.
The expanding complexity of wake models has introduced additional uncertain parameters beyond the wake spreading rate that must be prescribed.

An approach to automate the estimation of the unknown wake model parameters is to use Bayesian calibration, wherein the parameters are treated as uncertain variables to be calibrated---with quantified uncertainty---based on data \cite{kennedy_bayesian_2001}.
Bayesian parameter estimates have been thoroughly applied to the turbulence models of higher-fidelity computational fluid dynamics simulations \cite{cheung_bayesian_2011, edeling_bayesian_2014, ray_bayesian_2016}.
\citet{zhang_quantification_2020} (2020) used Markov chain Monte Carlo (MCMC) techniques to study the parameter uncertainty of a wake model in FLORIS, a controls-oriented wake modeling software package \cite{nrel_floris_2022}, and demonstrated improved predictive capabilities for mean velocity and power fluctuations. 
\citet{van_beek_sensitivity_2021} (2021) investigated this parameter uncertainty and sensitivity in FLORIS as well and found that the wake model is overparameterized.
Several studies have explored yaw angle optimization under wind condition and model parameter uncertainty---which requires a wake model with quantified uncertainty---and have found improved expected performance of the optimization results \cite{rott_robust_2018, simley_design_2020, quick_wake_2020, howland_wind_2021, van_beek_sensitivity_2021}.

For wake interactions from multiple turbines, individual wake velocity deficits are superimposed. 
Popular low-order superposition methods, such as those proposed by \citet{lissaman_energy_1979} and \citet{katic_simple_1986}, satisfy continuity but are not derived from first principles to include conservation of momentum.
Regardless, they are the standard methods in existing wake model softwares, such as FLORIS, and are often used in the literature as a result of their empirical predictive success \cite{sinner_power_2021}. 
Recently, \citet{zong_momentum-conserving_2020} (2020) derived a wake superposition method that conserves momentum. 
As a parallel approach, \citet{bastankhah_analytical_2021} (2021) recently developed a method that includes wake superposition in the wake velocity deficit model itself, avoiding the need for an independent wake superposition model.
\citet{howland_influence_2020} (2020) demonstrated that the superposition method influences the wake model parameter calibration and predictive success.
% Model error is expected from these simplified wake superposition models in addition to the low-fidelity wake deficit models with which they are used.
Key questions remain about the uncertainty of the parameter estimates and the accuracy of different superposition methods for predicting wake velocity. 
% As one component of this low-fidelity modeling framework, wake superposition methods incorporate critical---but still unknown---aspects of model error.
% In this study, we investigate the impact of a wider selection of superposition methods on the wake model parameter estimates and quantify their uncertainty.

There are two main objectives for this paper.
First, we interrogate the impact of superposition method on the wake model parameter estimates and model error, an effect which has previously been neglected, using an LES dataset.
Second, we test the predictive capabilities of this data-driven approach in a separate wake steering control application based on field data.
These two numerical experiments leverage Bayesian calibration of a Gaussian wake model with MCMC to quantify uncertainty.
For the second objective, we compare the data-driven predictions of turbine power production out of the training data sample (i.e. in a different flow scenario than the one for which the model parameters were calibrated) to those of the standard empirically-tuned estimates of the wake expansion rate.
With an improved understanding of the uncertainty associated with the wake deficit and superposition models, more accurate power predictions that are robust under uncertainty can inform control systems of wind farms and optimization studies of wind plant layouts \cite{van_beek_sensitivity_2021, howland_wind_2021}.

In Section \ref{sec:methods}, the velocity deficit model and superposition methods are outlined, the baseline estimate of the wake expansion rate is presented, and the Bayesian estimation framework is described. Section \ref{sec:setup} details the setup for the numerical experiments and the experimental data. Section \ref{sec:results} presents the results, including trends in the wake expansion coefficient, differences between five superposition methods, and out-of-sample power predictions.
Conclusions are provided in Section \ref{sec:conc}.

\section{Methods}\label{sec:methods}
\subsection{Wake Deficit and Superposition Models}\label{sec:models}

The velocity deficit downwind of a wind turbine is modeled with a two-dimensional Gaussian shape that includes yaw deflection \cite{bastankhah_experimental_2016}:
\begin{equation}\label{eq:gauss}
    \frac{\Delta u_{i,j}}{u_0} = \Bigg(1 - \sqrt{1 - \frac{C_T\cos{\gamma}}{8\sigma_y\sigma_z}}\Bigg) \exp{\Bigg(-\frac{(y-\delta)^2}{2\sigma_y^2} -\frac{ z^2}{2\sigma_z^2}\Bigg)},
\end{equation}
where $\Delta u_{i,j}$ is the wake deficit at turbine location $i$ produced by a turbine $j$ and relative to the incident velocity $u_0$. 
The streamwise, spanwise, and vertical coordinates are $x$, $y$, and $z$, respectively, and are provided relative to the upwind turbine's hub location and normalized by the rotor diameter $D$.
The yaw angle $\gamma$ is defined relative to the free-stream wind direction to be positive for counter-clockwise rotation about a vertical axis; yaw misalignment modifies the coefficient of thrust of the turbine ($C_T\cos{\gamma}$) and produces a spanwise wake deflection $\delta$.
The wake widths in the spanwise and vertical directions, $\sigma_y$ and $\sigma_z$, respectively, are defined by a linear growth with downstream position:
\begin{eqnarray}
    \sigma_y &= k(x-x_0) + \sigma_0 \cos\gamma, \\
    \sigma_z &= k(x-x_0) + \sigma_0,
\end{eqnarray}
where $k$ is the uncertain wake expansion rate.
The potential core length $x_0$, normalized by the rotor diameter, is a function of the yaw angle and thrust coefficient:
\begin{eqnarray}\label{eq:core}
    x_0 = \frac{\cos{\gamma}\:(1+\sqrt{1-C_T})}{\sqrt{2}(2.32I_0 + 0.154(1-\sqrt{1-C_T}))},
\end{eqnarray}
where the coefficients in the denominator (2.32 and 0.154) were calibrated by \citet{bastankhah_experimental_2016} (2016).
The initial wake width $\sigma_0$ is a function of $C_T$ \cite{bastankhah_new_2014}:
\begin{eqnarray}
    \sigma_0 = 0.2\sqrt{\frac{1 + \sqrt{1 - C_T}}{2\sqrt{1 - C_T}}},
\end{eqnarray}
with an empirically calibrated coefficient of 0.2 from \citet{bastankhah_new_2014} (2014).
We use these values of the empirical coefficients instead of calibrating them in this study to avoid overparameterizing the wake model.
Since the uncertainty in $k$, $x_0$, and $\sigma_0$ all contribute to the wake width and velocity deficit, it would be difficult to separate the compounded uncertainty of these unknown variables without more data beyond the power production of each turbine.
Consequently, $k$ is the only uncertain calibration parameter in this study; incorporating more data to calibrate more wake model parameters is suggested as an area of future work.

The centerline deflection $\delta$ is defined as follows:
\begin{eqnarray}\label{eq:deflection}
    \delta = \theta_{c0}x_0 + \frac{\theta_{c0}}{14.7}\sqrt{\frac{\cos{\gamma}}{k^2C_T}}(2.9+1.3\sqrt{1-C_T}-C_T) \\ \times \text{ln}\Bigg[\frac{(1.6+\sqrt{C_T})(1.6\sqrt{\frac{8\sigma_y\sigma_z}{\cos{\gamma}}}-\sqrt{C_T})}{(1.6-\sqrt{C_T})(1.6\sqrt{\frac{8\sigma_y\sigma_z}{\cos{\gamma}}}+\sqrt{C_T})}\Bigg],
\end{eqnarray}
where $\theta_{c0}$ is calculated by:
\begin{eqnarray}
    \theta_{c0} = \frac{-0.3\gamma}{\cos{\gamma}}(1-\sqrt{1-C_T\cos{\gamma}}).
\end{eqnarray}

The wake velocity is the difference between the horizontally homogeneous free-stream velocity $u_\infty$ and the velocity deficit, $u_{i,j} = u_\infty - \Delta u_{i,j}$. When more than one wake is incident to a given location, the wake velocity is calculated from the superposition of the individual deficits. We consider five different superposition methods in this paper:

\begin{itemize}
    \item \textbf{Lissaman \cite{lissaman_energy_1979}:} The wake velocity is defined relative to the freestream velocity $u_\infty$, and the wakes (from turbines $j$ at turbine location $i$) are superimposed linearly. The physical justification is analogous to pollutant dispersion in the atmospheric boundary layer, which can be superimposed linearly:
        \begin{eqnarray}
            u_i = u_\infty - \sum_{j}(u_\infty - u_{i,j}) = u_\infty - \sum_{j}u_\infty\bigg(\frac{\Delta u_{i,j}}{u_\infty}\bigg).
        \end{eqnarray}
    \item \textbf{Kati\'{c} \cite{katic_simple_1986}:} The wake velocity is defined relative to the freestream velocity $u_\infty$ and the wakes are superimposed as the sum of squares. The motivation is to superimpose kinetic energy deficits rather than velocity deficits:
        \begin{eqnarray}
            u_i = u_\infty - \sqrt{\sum_{j}(u_\infty - u_{i,j})^2} = u_\infty - \sqrt{\sum_{j}\bigg[u_\infty\bigg(\frac{\Delta u_{i,j}}{u_\infty}\bigg)\bigg]^2}.
        \end{eqnarray}
    \item \textbf{Niayifar \cite{niayifar_analytical_2016}:} The wake velocity is defined relative to the inflow velocity at turbine $j$, $u_j$, with a linear superposition. Lissaman's method has been found empirically to overestimate the velocity deficit after several rows of turbines \cite{crespo_survey_1999}, which this method aims to correct:
        \begin{eqnarray}
            u_i = u_\infty - \sum_{j}(u_j - u_{i,j}) = u_\infty - \sum_{j}u_j\bigg(\frac{\Delta u_{i,j}}{u_j}\bigg).
        \end{eqnarray}
    \item \textbf{Voutsinas \cite{voutsinas_analysis_1990}:} The wake velocity is defined relative to the velocity at turbine $j$ and with the sum-of-squares superposition. The justification for this method is both the superposition of energy deficits and the definition of the wake relative to local conditions:
        \begin{eqnarray}
            u_i = u_\infty - \sqrt{\sum_{j}(u_j - u_{i,j})^2} = u_\infty - \sqrt{\sum_{j}\bigg[u_j\bigg(\frac{\Delta u_{i,j}}{u_j}\bigg)\bigg]^2}.
        \end{eqnarray}
    \item \textbf{Zong \cite{zong_momentum-conserving_2020}:} Assuming steady, spatially-uniform inflow and neglecting turbulent transport, streamwise momentum is conserved by defining a mean convective wake velocity of the wake produced by turbine $j$, $u_{i,j}^{(c)}$:
        \begin{eqnarray}
            u_{i,j}^{(c)} = \frac{\int\int_A u_{i,j}\: \Delta u_{i,j}\:dA}{\int \int_A \Delta u_{i,j}\:dA}.
        \end{eqnarray}
    Here, the double integral is performed over the wake cross-section $A$. The total wake deficit at a location $i$ is computed as the weighted sum of the individual deficits, where the weight is the wake's proportion of the total mean convective wake velocity, $u_{i}^{(c)}$:
        \begin{eqnarray}
            \Delta u_{i} = \sum_j \frac{u_{i,j}^{(c)}}{u_{i}^{(c)}}\Delta u_{i,j}.
        \end{eqnarray}
    The total wake velocity deficit is used to calculate a new total mean convective wake velocity in an iterative fashion until $u_{i}^{(c)}$ converges:
        \begin{eqnarray}
            u_{i}^{(c)} = \frac{\int\int_A u_{i}\: \Delta u_{i} \:dA}{\int \int_A \Delta u_{i}\:dA}.
        \end{eqnarray}
    The wake velocity $u_{i}$ that results from this converged solution is the total wake velocity at the location $i$:
        \begin{eqnarray}
            u_i = u_\infty - \Delta u_{i}.
        \end{eqnarray}
\end{itemize}

We focus on Region II wind turbine operation with a constant coefficient of power $C_P$ and we prescribe constant density $\rho$.
Under yawed conditions, the coefficient of power is reduced by a factor $\cos^p(\gamma_i)$, where $p$ is an empirical constant, to be prescribed in later sections specific to the turbine data.
All wind turbines considered have the same rotor swept area $A_0$.
The turbine power is normalized relative to the upstream (free-stream) turbine under yaw alignment:
\begin{align}
    P_i &= \cfrac{\frac{1}{2} C_{P} \cos^p(\gamma_i)\rho A_0 u_i^3}{\frac{1}{2} C_{P} \cos^p(\gamma_1=0) \rho A_0 u_\infty^3}, \\ 
    P_i &= \cos^p(\gamma_i)\Big(\frac{u_i}{u_\infty}\Big)^3.
\end{align}

\subsection{Empirical Tuning Parameter Estimation Baseline}\label{sec:empirical}

Often, $k$ is estimated as a function of streamwise turbulence intensity \cite{lissaman_energy_1979} ($I$). 
An empirical relationship between $k$ and $I$ was proposed by \citet{niayifar_analytical_2016} (2016) for neutral conditions ($0.065 < I < 0.15$),

\begin{equation}\label{eq:k_ti}
k = 0.384I + 0.004.
\end{equation}
For Eq.~\ref{eq:k_ti} to be predictive, a model is also required for the turbulence intensity in the wind farm.
The turbulence intensity has two components---ambient ABL and wake-added. Crespo and Hern\'{a}ndez proposed an empirical model for wake-added turbulence \cite{crespo_turbulence_1996}:
\begin{eqnarray}\label{eq:prediction}
    I_{+,j} = 0.73\Bigg(\frac{1-\sqrt{1-C_T}}{2}\Bigg)^{0.8325} I_0^{0.0325} x^{-0.32},
\end{eqnarray}
where $I_0$ is the ambient turbulence intensity.
These individual wake-added turbulence contributions and the ambient turbulence are combined as a sum-of-squares \cite{crespo_turbulence_1996}:
\begin{eqnarray}\label{eq:TI_sos}
    I = \sqrt{I_0^2 + \sum_j{I_{+,j}^2}}.
\end{eqnarray}

As discussed by \citet{niayifar_analytical_2016} (2016), the empirical relationship (Eq.~\ref{eq:k_ti}) was found using idealized neutral conditions and neglecting stability.
The empirical correlation between $k$ and $I$ is sensitive to the atmospheric conditions, the operating conditions, and the turbines in question \cite{howland_wind_2019, doekemeijer_closed-loop_2020, howland_optimal_2020}.
Despite this limitation, the empirical relationship is the primary functional form used to estimate $k$ in wake modeling applications such as FLORIS and PyWake \cite{pedersen_dtuwindenergypywake_2019}. 
In this study, we compare data-driven parameter estimates to the baseline predictions of Eq.~\ref{eq:k_ti}.

\subsection{Data-Driven Parameter Estimation}\label{sec:bayesian}
\subsubsection{Bayesian Problem Definition}
In this study, we develop a Bayesian calibration approach to estimate the wake model parameters $\mathbf{k}$ for each turbine and their uncertainty.
This Bayesian calibration approach is an alternative to the empirical tuning method introduced in Section \ref{sec:empirical}.
We consider a general relationship between a model and an observation as follows \cite{kennedy_bayesian_2001}:
\begin{eqnarray}
    \mathbf{y} = \boldsymbol{\zeta} (\boldsymbol{\theta}, \mathbf{x}) + \mathbf{e} = \mathbf{g} (\boldsymbol{\theta}, \mathbf{x}) + \boldsymbol{\delta} (\boldsymbol{\theta}, \mathbf{x}) + \mathbf{e}.
    \label{eq:inverse}
\end{eqnarray}

In this relationship, an observation $\mathbf{y}$ is the sum of a true process $\boldsymbol{\zeta}$ and an observation uncertainty $\mathbf{e}$.
The process $\boldsymbol{\zeta}$ depends on calibration parameters $\boldsymbol{\theta}$ and some other known model inputs $\mathbf{x}$, and is approximated with a model $\mathbf{g}$, which has model error $\boldsymbol{\delta}$.  

In this study, $\mathbf{y}$ is an individual observation of the finite-time average power production of each turbine in the wind farm in the set $\mathcal{Y}$: 
\begin{eqnarray}
    \mathcal{Y} = \{\mathbf{y_1},..., \mathbf{y_m},...\},
\end{eqnarray}
where $\mathbf{y_m} = [P_{1,m},P_{2,m},...]^\mathbf{T}$ with one element for each turbine.
The output of our wake model, $\mathbf{g}$, is a prediction of this power production.
The calibration parameters in this study are the unique wake expansion rate $k$ for each turbine in the wind farm, and all other model inputs are known:
\begin{eqnarray}
    \boldsymbol{\theta} = [k_1, k_2, \cdots], \\
    \mathbf{x} = [x,y,z,z_H,D,C_T,u_{\infty,H},\phi_{\infty,H}],
\end{eqnarray}
where $z_H$ is the turbine hub height.
The wind direction $\phi_{\infty,H}$ and wind speed $u_{\infty,H}$ are assumed to be uniform in space and height (i.e. no wind veer or shear).

The characterization of the wake model error $\boldsymbol{\delta}$ is difficult because, if it were known, the model would be corrected to account for it.
Therefore, similar to other studies using Bayesian inference for assessing model parameters \cite{ray_bayesian_2016, zhang_quantification_2020}, we neglect the unknown wake model error.
This assumption yields:
\begin{eqnarray}
    \mathbf{y} = \mathbf{g} (\boldsymbol{\theta}, \mathbf{x}) + \mathbf{e}.\label{eq:inverse2}
\end{eqnarray} 
Here, variability in the observations $\mathbf{y}$ is explained by the parameter uncertainty in $\boldsymbol{\theta}$ and the observation uncertainty $\mathbf{e}$.
By neglecting $\boldsymbol{\delta}$, wake velocity deficit and superposition model errors will be reflected in the estimation process and uncertainty quantification of $\boldsymbol{\theta}$.
As a result, differences between parameter estimates for various wake model superposition methods will partially elucidate the role of superposition modeling error.

The observational uncertainty $\mathbf{e}$ results from variability in the observed power production due to turbulence: the mean power production is computed with a finite-time average that only converges to a deterministic value with an infinite time-averaging length.
We make two assumptions about this uncertainty.
First, we assume zero bias. 
Along with our assumption of zero modeling error, we therefore are assuming that the wake model accurately predicts the mean of the observed data in the limit of many observations (or an infinite time-average).
Second, we assume the distribution of each turbine's power production is Normal according to the central limit theorem: $\mathbf{e} \sim \mathcal{N}(0,\Sigma)$.
The observational covariance matrix $\boldsymbol{\Sigma}$ parameterizes the uncertainty of the observations, where the element $\Sigma_{ij}$ is the covariance of the power production of two turbines $i$ and $j$:
\begin{eqnarray}
    \Sigma_{ij} = \frac{1}{M-1}\sum_{m=1}^M (P_{i,m} - \overline{P_i})(P_{j,m} - \overline{P_j}).
\end{eqnarray}
For field measurements, observational uncertainty also includes variability due to non-stationary wind conditions (i.e. mean wind speed, mean wind direction, atmospheric stability) and measurement error. 

% this assumption is valid only if the system is statistically stationary and sufficiently many time-averaged measurements are taken.

% Since the observational error is assumed to be Normally-distributed, and the model is deterministic, then the distribution for an individual observation $\mathbf{y}_m$ is also assumed to be Normally-distributed:
% \begin{eqnarray}
%     (\mathbf{y}_m|\boldsymbol{\theta}) \sim \mathcal{N}(\boldsymbol{\mu},\boldsymbol{\Sigma}),
% \end{eqnarray}
% where $\boldsymbol{\mu}$ is the model output $\mathbf{g}(\boldsymbol{\theta}, \mathbf{x})$ because the observational uncertainty has zero bias. 

\subsubsection{MCMC Formulation and Algorithm}
Bayesian inference is based on Bayes' rule: $\mathcal{P}(\boldsymbol{\theta}|\mathcal{Y}) \propto \mathcal{L}(\mathcal{Y}|\mathbf{\boldsymbol{\theta}}) \Pi(\boldsymbol{\theta})$. Given some information about the possible values of the calibration parameters (i.e. the prior distribution $\Pi(\boldsymbol{\theta})$), and a function that describes the conditional probability of the observations given those parameters ($\mathcal{L}(\mathcal{Y}|\boldsymbol{\theta})$), we can estimate the posterior distribution $\mathcal{P}(\boldsymbol{\theta}|\mathcal{Y})$, the conditional probability of the parameters given the observations and the prior.

The objective function for an individual observation is the difference between the observation and the model output: $\boldsymbol{\epsilon}_m = \mathbf{y}_m - \mathbf{g}(\boldsymbol{\theta}, \mathbf{x})$.
The likelihood function for this individual observation, based on the Normally-distributed observational uncertainty, is
\begin{eqnarray}\label{eq:likelihood}
    \mathcal{L}_m(\mathbf{y}_m|\boldsymbol{\theta}) \propto \frac{1}{\sqrt{|\boldsymbol{\Sigma}|}}\exp\Big(-\frac{1}{2}\boldsymbol{\epsilon}_m^{\mathbf{T}}\boldsymbol{\Sigma}^{-1}\boldsymbol{\epsilon}_m\Big).
\end{eqnarray}
The total likelihood function across all observations is the product of the likelihood function of each observation:
\begin{eqnarray}\label{eq:likelihood_total}
    \mathcal{L}(\mathcal{Y}|\boldsymbol{\theta}) \propto \prod_m\frac{1}{\sqrt{|\boldsymbol{\Sigma}|}}\exp\Big(-\frac{1}{2}\boldsymbol{\epsilon}_m^{\mathbf{T}}\boldsymbol{\Sigma}^{-1}\boldsymbol{\epsilon}_m\Big),
\end{eqnarray}
assuming that the observations $\mathbf{y}_m$ are independent.

% \begin{equation}
%     \mathcal{P}(\boldsymbol{\theta}|\mathcal{Y}) \propto \Pi(\boldsymbol{\theta}) \prod_m\frac{1}{\sqrt{|\boldsymbol{\Sigma}|}}\exp\Big(-\frac{1}{2}\boldsymbol{\epsilon}_m^{\mathbf{T}}\boldsymbol{\Sigma}^{-1}\boldsymbol{\epsilon}_m\Big).
% \end{equation}

In order for the velocity deficit (Eq.~\ref{eq:gauss}) to be real-valued, we require a lower limit on $k$:
\begin{equation}\label{eq:limit}
    k \geq \frac{\sqrt{\frac{1}{8}C_T\cos{\gamma}} - \sigma_0}{x}.
\end{equation}
In practice, we wish to sample values of $k$ (i.e. $\boldsymbol{\theta})$ in an unconstrained space to simplify the proposal of new candidate parameter values. We use the bound in Eq.~\ref{eq:limit} as a lower limit $k_{low}$, choose a finite but sufficiently-large upper bound $k_{high}$, and transform $k$ from physical to transform space using the following equations:
\begin{align}\label{eq:transform}
    k_{T} &= \text{ln}\Bigg[\cfrac{k - k_{low}}{k_{high} - k}\Bigg], \\
    k &= \cfrac{k_{high} \exp(k_{T}) + k_{low}}{\exp(k_{T}) + 1},
\end{align}
where $k_T$ is the parameter value in the unconstrained transform space.

We define the prior distribution $\Pi(\boldsymbol{\theta_T})$ in the transform space with a Normal distribution: $\Pi(k_T) \sim \mathcal{N}(k_{T_0}, \sigma_\pi^2)$; non-Gaussian posteriors are permitted despite the use of the Gaussian prior.
The prior mean $k_{T_0}$ is set at the transformed value of $k$ using the empirical relation in Eq.~\ref{eq:limit} with the ambient turbulence intensity. The standard deviation of the prior is prescribed to be sufficiently high ($\sigma_\pi = 2.5$, i.e. the prior is uninformative outside of the physical constraints) such that the posterior distribution is not sensitive to it, based on numerical experiments not shown.

To compute the posterior distribution, MCMC is applied by implementing the random-walk Metropolis algorithm \cite{brooks_handbook_2011}. We initialize the calibration parameters in the transform space with the mean values of the prior, $\boldsymbol{\theta_T^{(0)}} = \mathbf{k_{T_0}}$.
At iteration $n$, a proposal set of the parameters, $\boldsymbol{\theta_T^*}$, is generated from a symmetric proposal distribution centered on the parameter set from the previous iteration, $\boldsymbol{\theta_T^{(n-1)}}$:
\begin{eqnarray}
    \boldsymbol{\theta_T^*} = \boldsymbol{\theta_T^{(n-1)}} + \eta\:\mathcal{N}(0,\sigma_\pi^2).
\end{eqnarray}
The proposal distribution for $k$ is independent for each turbine. 
The proposal step factor $\eta$ is a prescribed parameter. 
The parameter values are then transformed back into physical space ($\mathbf{k_T}$ to $\mathbf{k}$) and input to the wake model.
Our acceptance criteria is based on the ratio of the posterior probability of the proposed parameters to that of the previous parameters,
\begin{eqnarray}
    \alpha = \frac{\mathcal{L}(\mathcal{Y}|\boldsymbol{\theta^*}) \Pi(\boldsymbol{\theta_T^*})}{\mathcal{L}(\mathcal{Y}|\boldsymbol{\theta^{(n-1)}}) \Pi(\boldsymbol{\theta_T^{(n-1)}})},
\end{eqnarray}
and is defined as follows: if a uniform random variable $c \in [0,1] \leq \alpha$, then accept the candidate such that $\boldsymbol{\theta^{(n)}} = \boldsymbol{\theta^*}$; if not, then reject the candidate such that $\boldsymbol{\theta^{(n)}} = \boldsymbol{\theta^{(n-1)}}$.

The step factor $\eta$ is tuned such that the global acceptance rate is around 20\%, which is widely considered to be optimal \cite{gelman_weak_1997}. A burn-in period $\tilde{N}$ is used to reduce the effect of the initial parameters $\boldsymbol{\theta^{(0)}}$, which we set at $\tilde{N}=1000$ based on \citet{kruschke_doing_2014} (2014). Finally, since the samples are correlated due to the Markov chain nature of the algorithm, every fifth posterior sample is saved.

\section{Numerical Experiment Setup}\label{sec:setup}

\begin{figure*}
    \centering
    \includegraphics[width=0.99\textwidth]{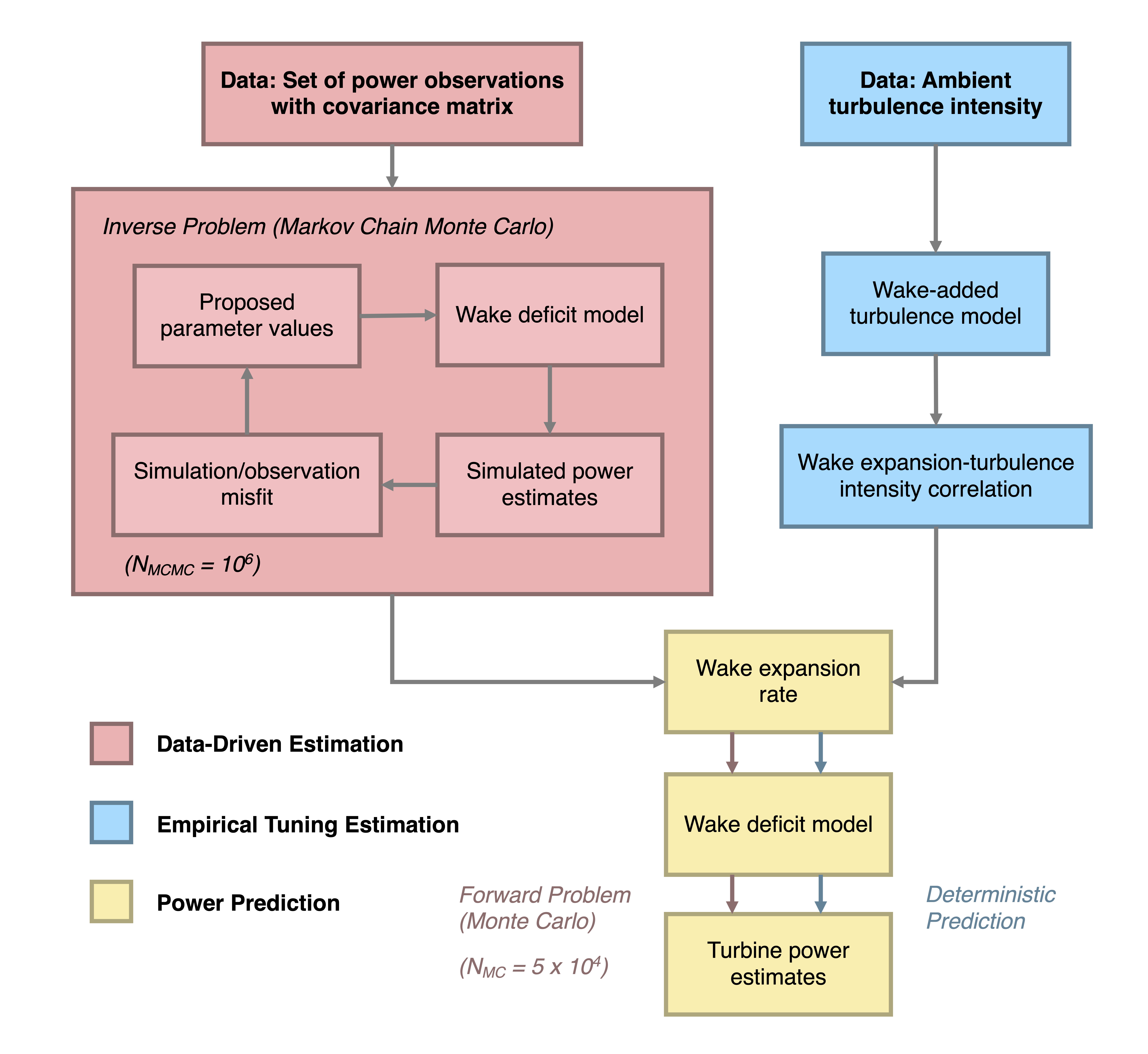}
    \caption{A flowchart illustrating the interaction between different modeling procedures in the power estimation process, where bold text denotes inputs. The ``forward problem'' using Monte Carlo refers to power prediction with the data-driven estimates only.
    \label{fig:flowchart}
    }
\end{figure*}

Fig. \ref{fig:flowchart} illustrates our modeling framework. 
We use the wake model described in Section \ref{sec:models} to estimate turbine power production. 
We term this the ``forward'' problem: inputting wake model parameters $\mathbf{k}$ to the wake deficit model and outputting turbine power. 
One way to produce values for the wake expansion rate is through the empirical tuning method described in Section \ref{sec:empirical}. 
This approach outputs a single value of $k$ for each turbine.

In Section~\ref{sec:bayesian}, we developed a second, data-driven approach for estimating the wake model parameters. 
This framework requires training data in the form of the set of observations of power production for all turbines in the wind farm and outputs posterior distributions of the wake expansion rate for each turbine. 
We refer to this as the ``inverse'' problem. 
The parameter values resulting from the output distributions from MCMC can be randomly sampled and fed into the wake model to estimate a probability distribution of wind turbine power (i.e. Monte Carlo forward problem).
These power predictions are then compared to a separate test dataset (or to the training data in a perfect-model experiment).
In our experiments, we perform one million MCMC simulations ($N_{MCMC} = 10^6$) and fifty thousand Monte Carlo simulations ($N_{MC} = 5\times10^4$), which are sufficiently large to ensure convergence \cite{kruschke_doing_2014} without unnecessary computational expense.

\subsection{Large Eddy Simulation Dataset}\label{sec:exp1}

\begin{figure}
    \centering
    \includegraphics[width=0.5\textwidth]{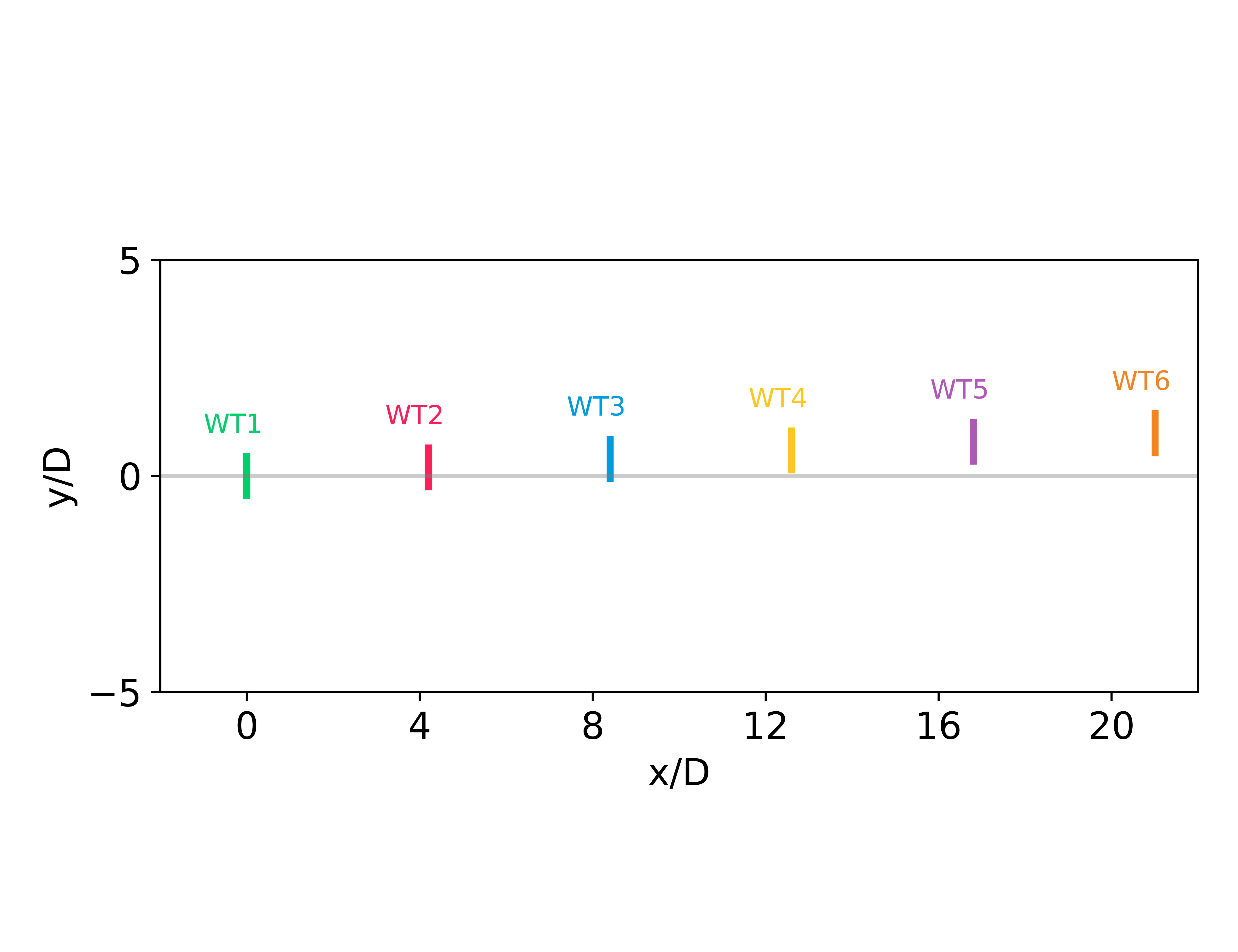}
    \caption{Turbine layout in the large eddy simulation with the wind aligned with the positive x-direction.
    \label{fig:farm1}
    }
\end{figure}

The purpose of this first numerical experiment is to explore the sensitivity in the posteriors of $k$ to the choice of superposition method as a way of illustrating modeling error.
We consider a six turbine wind farm, as shown in Fig. \ref{fig:farm1}. 
We will refer to each turbine with `WT' followed by an index in ascending order from upstream to downstream.
The turbine spacing is approximately four rotor diameters in the streamwise direction, and the turbines are roughly aligned with respect to the incoming wind direction at hub-height.
The spacing between the wind turbines was selected to approximately represent the commercial wind farm of interest in \citet{howland_wind_2019} (2019). The rotor diameter of all turbines is $D = 126$ m, hub height is $z_H = 100$ m, the wind speed at hub height is $u_{\infty,H} = 8$ m/s, the thrust coefficient is $C_T = 0.75$, and the ambient turbulence intensity is $I_0 = 7.7\%$.

The observation data is recorded from LES under conventionally neutral atmospheric stability. Power is sampled from each turbine at the timescale of the LES timestep and collated in 30-minute time-averages, with 50 independent samples.
The data sampling is initialized after the simulation has reached a statistically quasi-stationary state (see further details on conventionally neutral ABL statistical quasi-stationarity in \citet{allaerts_large_2015} and \citet{howland_influence_2020-1}).
We select LES data with fixed mean wind speed and direction to reduce modeling error by satisfying the horizontal homogeneity assumption in the derivation of the wake model and to support the assumption of Normally-distributed observations according to the central limit theorem.
We use the 30-minute averages because it is sufficiently larger than the large eddy turnover timescale; however, this finite averaging length is insufficiently long to remove all of the randomness due to turbulence.
As a result, observational error in the power production samples exists due to turbulent fluctuations.
Finally, in order to focus on the effect of superposition on the wake model parameter estimates and its contribution to modeling error, we use an idealized setup by estimating the wake model parameters with the same training dataset with which we compare the power predictions in the forward problem.
A more principled test of predictive accuracy against hold-out data is described in the next section.

\subsection{Utility-Scale Wind Farm Dataset}\label{sec:exp2}

\begin{figure}
    \centering
    \includegraphics[width=0.5\textwidth]{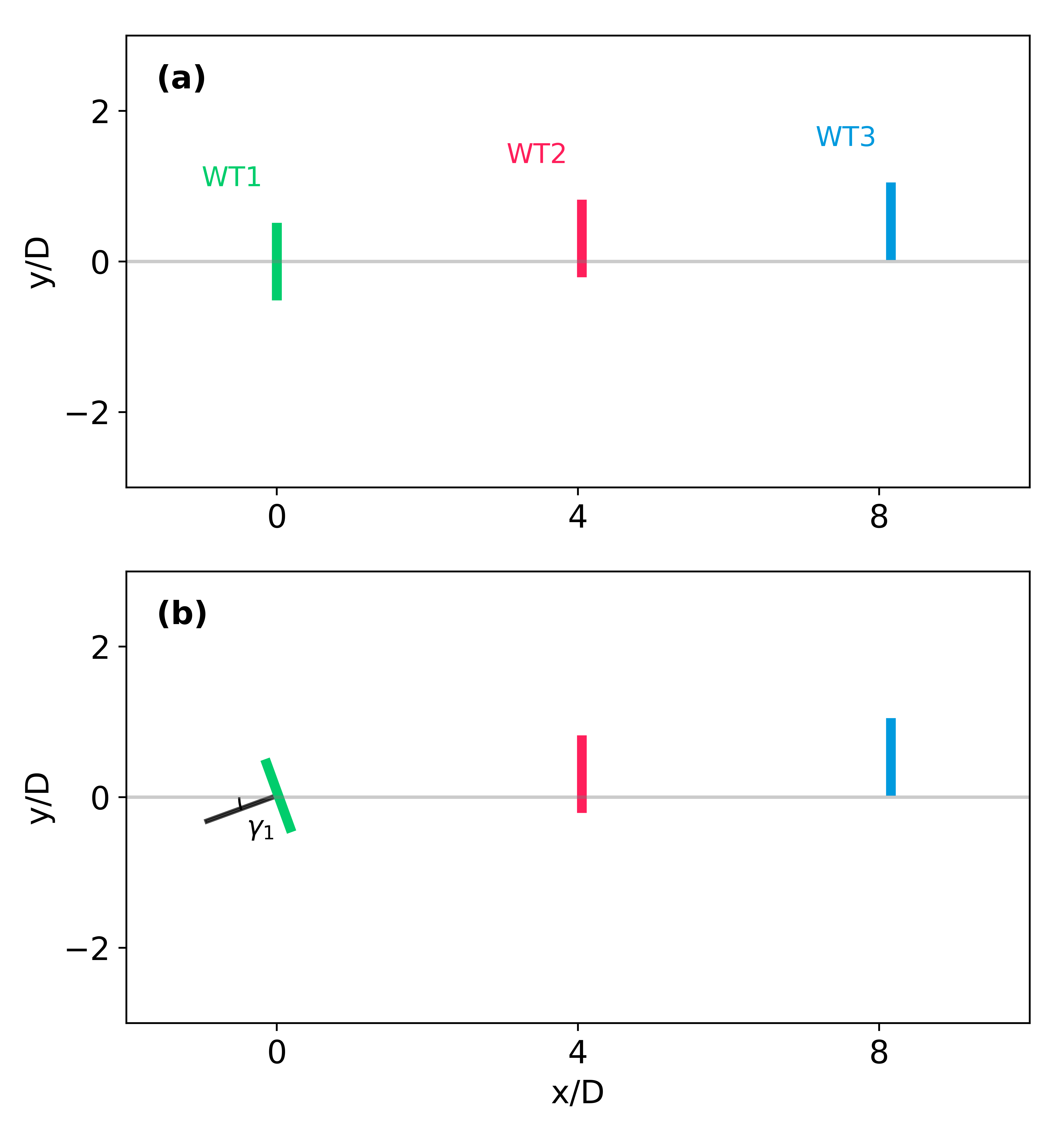}
    \caption{Turbine layout of the three turbines under consideration in the utility-scale wind farm, with the wind aligned with the positive x-direction. (a) The no-yaw (training) case used for calibration of the wake model parameters. (b) The yaw (test) case used for power prediction.
    \label{fig:farm2}
    } 
\end{figure}

The purpose of this second numerical experiment is to explore the predictive capabilities of the two methods for estimating the unknown wake expansion rate in a scenario outside of the calibration sample.
We consider a different wind farm with three turbines, as shown in Fig. \ref{fig:farm2}.
The turbine spacing is again roughly four rotor diameters in the streamwise direction with similar alignment with the incident wind direction.
This dataset represents the wind farm studied in \citet{howland_optimal_2022} (2022). 
The rotor diameter of each turbine is $D = 120$ m, hub height is $z_H = 100$ m, wind speed is $u_{\infty,H} = 7$ m/s, the thrust coefficient is $C_T = 0.83$, and the ambient turbulence intensity is $I_0 = 5.0\%$. 
The cosine power exponent $p = 2$ for this scenario.

Power is sampled from each turbine and collected in 10-minute time-averages, with 116 samples for the no yaw case and 84 samples for the yaw case.
The raw dataset consists of independent 1-minute time-averages sampled by each turbine's Supervisory Control and Data Acquisition (SCADA) system.
We generate distributions of 10-minute time-averaged power production by bootstrapping: ten independent 1-minute time-averaged power production samples are ensemble-averaged into a new power production distribution.
The physical justification for this longer time-average is that a 1-minute average is insufficient to resolve the mean flow.
The Gaussian wake model is a steady-state model, derived as a solution to simplified Reynolds-averaged Navier-Stokes equations \cite{bastankhah_experimental_2016}.
Therefore, for the observations to represent mean power production, the field data must be averaged over a sufficiently long period to contain the energetic turbulent timescales.
The Appendix contains further details of this analysis.

% The boundary conditions of this field dataset are not fixed as they were in the LES dataset.
% Therefore, uncertainty in the power predictions is caused by this uncertainty in the boundary conditions, in addition to the uncertainty due to the finite time-averaging length.
% We include all of this uncertainty in $\mathbf{e}$ and test the predictive accuracy of the wake models in this less-idealized operating environment.

Data are available for two operating scenarios: one where all turbines are aligned with the incident wind ($\boldsymbol{\gamma} = 0^\circ$), and a second where the upstream turbine (WT1) is yawed at $\gamma_1 = 20^\circ$. 
We assume that the wake expansion rate does not depend on the yaw angle of the turbine (which is also an assumption used by the turbulence intensity-based model of $k$ in Eq.~\ref{eq:k_ti}).
We perform the inverse problem to estimate the wake expansion rate using only the power production with no yaw as training data.
Hence, the resulting posteriors are not informed by the data with the yawed upstream turbine.
We then sample from these posteriors to provide a forward prediction for the problem where the upstream turbine is yawed, which reduces the power production of the upstream turbine and laterally deflects its wake.
For comparison, we also predict the turbine power production for the yawed problem with the empirical-tuning estimates for $k$.

\section{Results}\label{sec:results}
\subsection{LES Data Results}
\subsubsection{Inverse Problem}
\begin{figure*}
    \centering
    \includegraphics[width=0.99\textwidth]{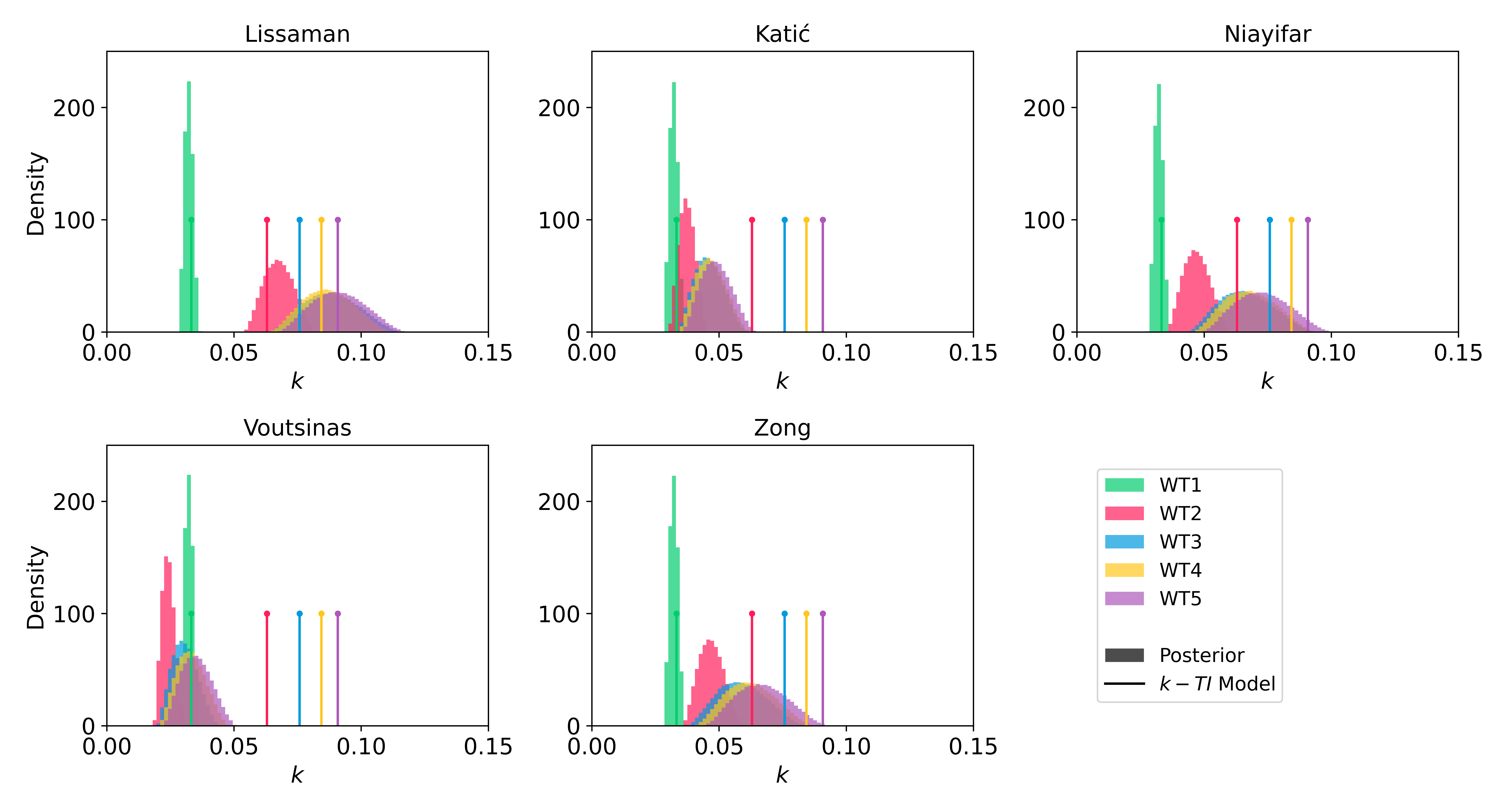}
    \caption{Estimates of the wake expansion rate $k$ using five different wake deficit superposition methods. The posteriors are generated using our data-driven approach, while the predictions from the $k-TI$ model represent the empirical tuning method and do not depend on the superposition method. `WT' refers to each wind turbine, in ascending order from upstream to downstream.
    \label{fig:superposition}
    } 
\end{figure*}

Fig.~\ref{fig:superposition} shows the posterior distributions for the wake model parameters for each turbine in the wind farm, obtained using the procedure outlined in Section~\ref{sec:setup}.
The figure includes five different results obtained by performing the inverse problem separately for each of the five superposition methods identified in Section~\ref{sec:models}.
The posterior distributions are compared to the deterministic values for $k$ obtained using the standard empirical tuning method.
Note that we do not consider the wake expansion rate of WT6 because it has no influence on the power production of any turbines in the farm.

The common trend among all but one of the estimates of the wake expansion rate is that $k$ increases as we move downstream in the wind farm.
Physically, the wake-added turbulence contribution from each turbine increases the wake recovery rate of the next turbine downwind.
The marginal increase in $k$ decreases with each additional turbine.
In an infinitely-long wind farm, a fully-developed region forms where the flow characteristics asymptotically approach a stationary value and there is no change between successive turbines \cite{calaf_large_2010}; by WT4, this trend becomes visible as the posteriors of $k$ begin to collapse on top of each other.
Although the Bayesian calibration permits non-Gaussian posteriors, the posteriors appear to be roughly Gaussian and the variance of $k$ grows for each successive turbine, which represents the compounding of uncertainty of the parameters for upstream turbines.
%Physically, it is possible that the wake-added turbulence introduced by the upstream turbines contributes variability and error in modeling the downstream wake expansions.
Note that the estimates for $k_1$ are roughly identical (with small differences due to randomness inherent to the MCMC method) regardless of choice of superposition method.
There is no superposition involved in the wake interaction between WT1 and WT2, so this insensitivity to superposition method is also expected physically.
The fact that these physical trends are captured by our data-driven estimation of $\mathbf{k}$ establishes confidence in our method.

Beneath this broad trend, there are qualitative differences between the five superposition methods stemming from their fundamental definitions.
Compared with the Lissaman model, the Kati\'{c} method produces posterior distributions with smaller means and smaller variances.
The smaller means can be explained by the sum-of-squares superposition allocating less weight (i.e. smaller magnitude) to the upstream turbine wakes compared to a linear method for a given set of input wake expansion rates since the normalized wake deficits are always less than one.
So, for fixed input $\mathbf{k}$, the sum-of-squares superposition predicts higher power production compared to the linear superposition.
However, since power production is the fixed input in this numerical experiment, by this reasoning the Kati\'{c} method requires that the values of $k$ are smaller to create a larger velocity deficit in order to predict the same power as the Lissaman method.
The variability is also diminished because the reduced weight assigned to the smaller wake deficits in the sum-of-squares results in a narrower range of wake velocities.

Niayifar, compared with Lissaman, also produces smaller mean values for $k$, but with similar variability.
The smaller means here are caused by the wake deficit being defined relative to a smaller reference velocity since the velocity within a wake is always lower than the freestream value.
So, for a given $\mathbf{k}$, the wake velocity (and by extension, power production) would be larger.
Following the same logic as the previous discussion, this instead leads to $k$ being smaller for this method compared with Lissaman.
However, the variability is not substantially changed because the linear superposition is still used.

The Zong and Niayifar posteriors are almost identical, with Zong tending to estimate slightly lower values of $k$ compared to Niayifar; Zong's model is equivalent to Niayifar in the limit of infinite separation between turbines (i.e. small wake velocity deficits) \cite{zong_momentum-conserving_2020}, so we would expect their posteriors to be similar.
As for the Voutsinas model, the anticipated trend of increasing $k$ with downstream position is absent. The variability of the posteriors is similar to Kati\'{c}, but a decrease in the wake expansion rate in the interior of the wind farm is unexpected.

Between these five methods, it is clear that the estimate of the wake expansion coefficient is highly sensitive to the choice of superposition method.
For example, the mean of $k$ for WT2-WT5 is roughly 80\% larger for Lissaman than Kati\'{c} and roughly 90\% larger for Zong than Voutsinas.
The variability also varies significantly, with differences in the standard deviation of $k$ up to 250\% between methods.

These differences in the posteriors indicate the presence of systematic model error in the superposition methods that would introduce discrepancies in turbine power predictions.
In theory, $k$ is supposed to represent the growth rate of the wake; in practice, it can be calibrated as a random model parameter to represent systematic model error. When doing the latter, it is of fundamental importance that the calibration and the prediction happens with the same combination of wake model and superposition model.

We also compare these posterior distributions to the empirical fit for $k$ based on turbulence intensity in Eq. \ref{eq:k_ti}.
The prediction follows the trend of $k$ increasing at a diminishing rate as we move downstream in the farm.
The Lissaman method appears to match most closely to the empirical tuning estimates compared with the other superposition methods; the posteriors of the Niayifar and Zong methods also contain the empirical tuning estimates.
On the other hand, the Kati\'{c} and Voutsinas posteriors underestimate $k$ by roughly 50\% compared to the empirical tuning method.
These results do not indicate that the empirical tuning method is a worse predictive tool than our data-driven method; we save our determination of predictive success for the field data experiment in Section \ref{sec:results2} where both methods are tested outside of their calibration sample.

\subsubsection{Forward Problem}
\begin{figure*}
    \centering
    \includegraphics[width=0.99\textwidth]{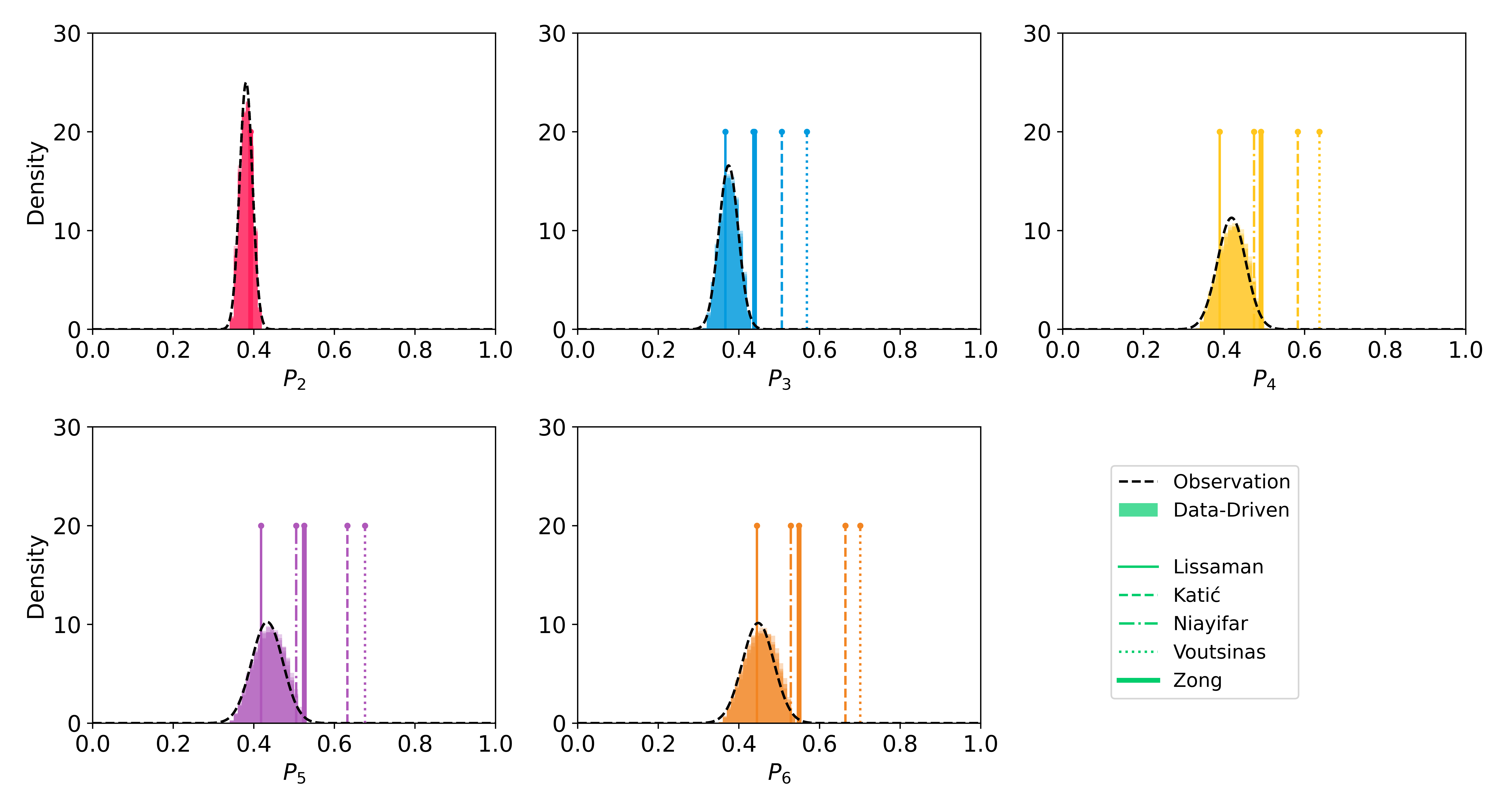}
    \caption{Power prediction for each turbine based on different estimates for the wake expansion rate. The data-driven approach to estimate $k$ for each of the five superposition methods produce distributions of power that all collapse to the observation (plotted as superimposed histograms). The empirical tuning estimate of $k$ produces different deterministic power estimates (plotted as lines) depending on the superposition method used.
    \label{fig:power1}
    } 
\end{figure*}

To verify the MCMC implementation and the posterior distributions they generated, we perform the forward problem as described in Section \ref{sec:setup}.
Fig. \ref{fig:power1} displays the results of the Monte Carlo power prediction simulations.
Note that we display the power production for all of the waked turbines WT2-WT6, since the normalized power production of WT1 is defined to be exactly unity.
The distributions of power prediction for all five superposition methods collapse onto the observed, Normally-distributed power production.
In this perfect-model experiment in which we calibrate with and compare to the same dataset, we expect the simulated power predictions to exactly match the observed power distributions for all superposition methods.
On the other hand, the predictions using the empirical tuning estimates of $k$ tend to overpredict turbine power since the estimated wake expansion rates are generally larger than those from the posteriors.

\begin{figure*}
    \centering
    \includegraphics[width=0.99\textwidth]{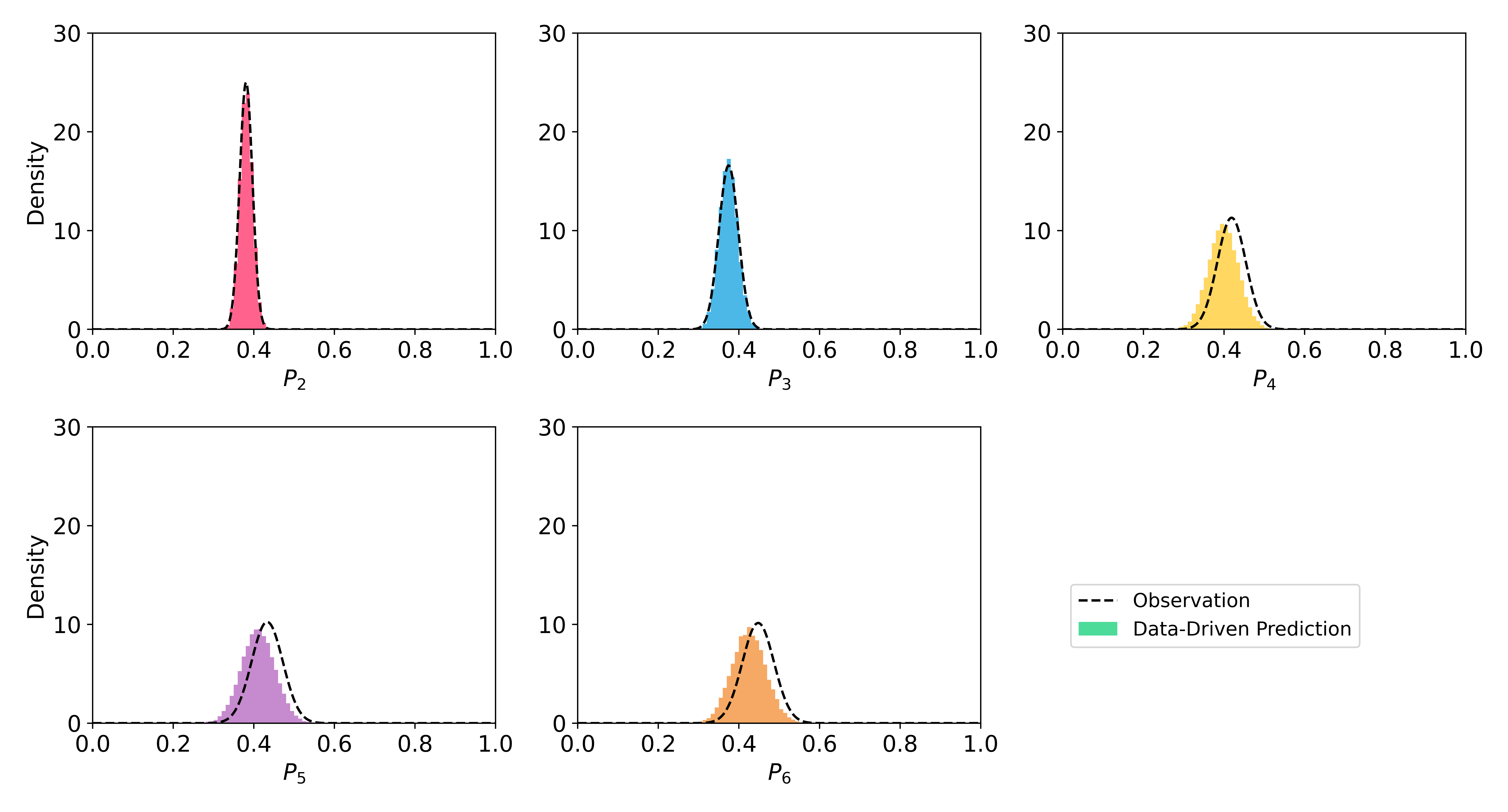}
    \caption{Power predictions using the Niayifar superposition method with wake expansion rates sampled from posteriors calibrated with the Zong method. Error in the mean power prediction for WT4-WT6 is about 5\%.
    \label{fig:error}
    } 
\end{figure*}

We consider one more scenario to reinforce the point that parameter calibration and power prediction should use the same superposition method.
From Fig. \ref{fig:superposition}, we found that the Niayifar and Zong posteriors were qualitatively very similar.
We can sample from the posterior distributions for $k$ that were calibrated using the Zong superposition method, but use the Niayifar superposition method for the power predictions in the forward problem.
The power predictions from this approach are shown in Fig. \ref{fig:error}.
We now observe a breakdown in the power predictions for WT4-WT6, with errors in the mean on the order of 5\%.
Across different combinations of superposition methods, the total error across the wind farm ranges from roughly 3\% in the best case to over 50\% in the worst case.

These results demonstrate that performing parameter estimation with one superposition method and power prediction with another method results in substantial errors.
Two parameters for the same physical phenomenon in two different models are not identical; in this context, we recognize that the wake expansion rate $k$ is not interchangeable between two different superposition models.
The wake combination method should be treated as a fixed component of the model, and calibration and prediction should be carried out with the same method.
Then, parameter estimation can be performed to tune the model to the operating conditions and make more accurate predictions.

This finding is cautionary for a modular approach to wake modeling.
Often, wake deficit models, superposition methods, and wake-added turbulence models are built as independent modules in steady-state wake modeling codes \cite{nrel_floris_2022, pedersen_dtuwindenergypywake_2019}.
Fig.~\ref{fig:error} illustrates that fixing the wake model parameters and swapping in a different superposition method could introduce modeling error depending on how the parameters were calibrated.
Future work could leverage this modular framework to continue exploring and quantifying wake superposition modeling error.

\subsection{Field Data Results}\label{sec:results2}
\subsubsection{Inverse Problem}
\begin{figure*}
    \centering
    \includegraphics[width=0.99\textwidth]{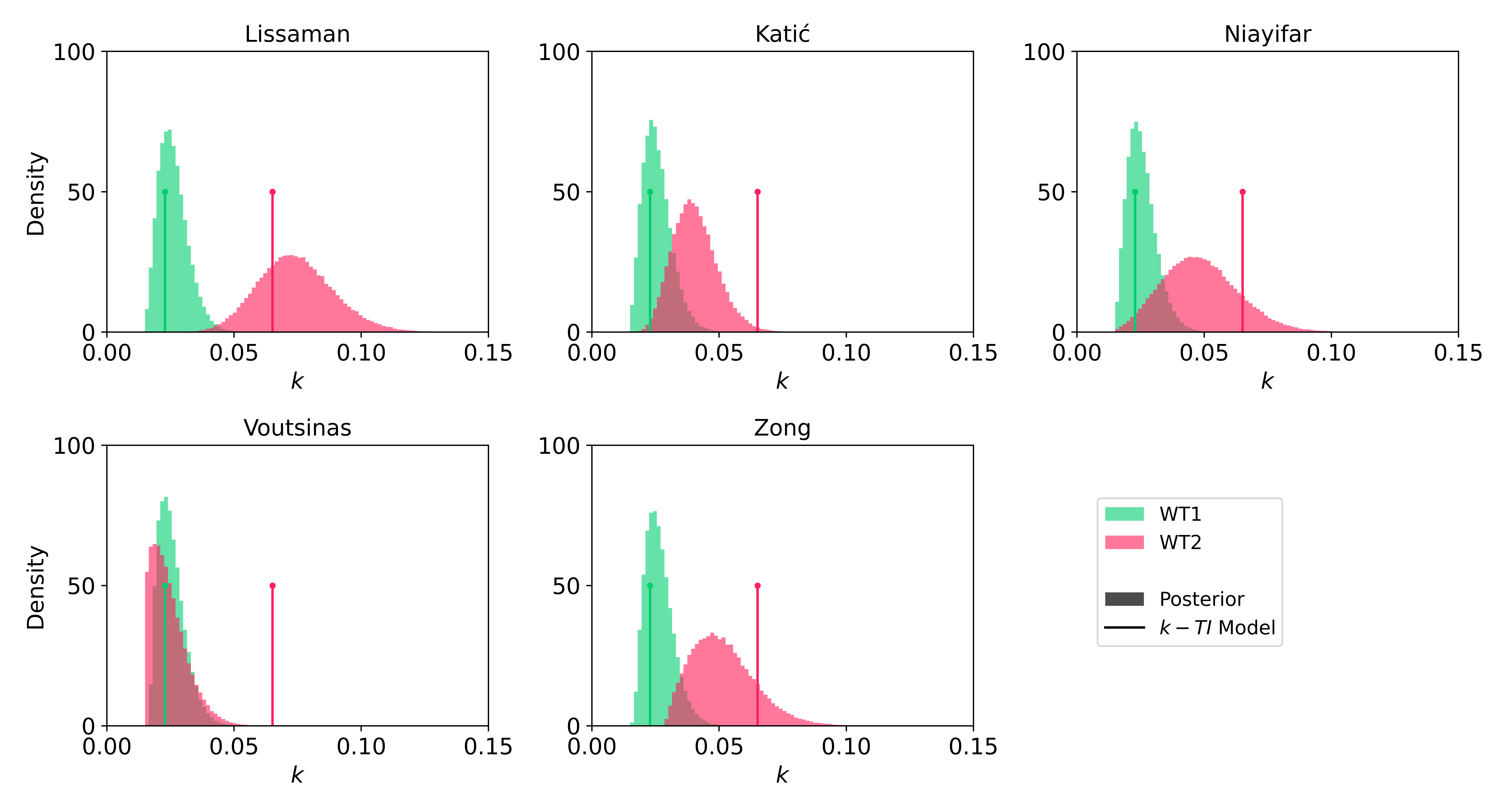}
    \caption{Wake expansion rate estimates for the field data with no yawed turbines using the data-driven and empirical tuning methods.
    \label{fig:posteriors}
    } 
\end{figure*}

Now we turn our attention to the field dataset, following the procedure outlined in Section \ref{sec:exp2}. 
The posteriors of $k$ for WT1 and WT2 from the wind farm in Fig.~\ref{fig:farm2}a are shown in Fig.~\ref{fig:posteriors}.
These results illustrate the same trends as in Fig~\ref{fig:superposition}, except that these posteriors possess more variance: as the observational uncertainty increased between the LES dataset and field dataset, the model uncertainty also increased (reflected in the uncertainty in $k$).
This result suggests that wake model predictions of real flows require accounting for larger modeling error due to the non-stationary atmospheric conditions.

% The skewness of the posteriors is also larger in this experiment.
% In the parameter transform space ($\boldsymbol{\theta_T}$), our posteriors are roughly Gaussian with small skewness.
% However, when the variability of the posteriors is high---as it is in this experiment---there is a substantial density of parameter values that map to a very narrow range of small magnitudes in physical space, given the logarithmic parameter transform (Eq.~\ref{eq:transform}).
% This nonlinear mapping produces the higher skewness in the posteriors in physical space.
% This effect was still present in the LES data test case (Fig.~\ref{fig:superposition}) but was not as visually obvious because of the reduced variability of the LES observations.

\subsubsection{Forward Problem}
\begin{figure*}
    \centering
    \includegraphics[width=0.99\textwidth]{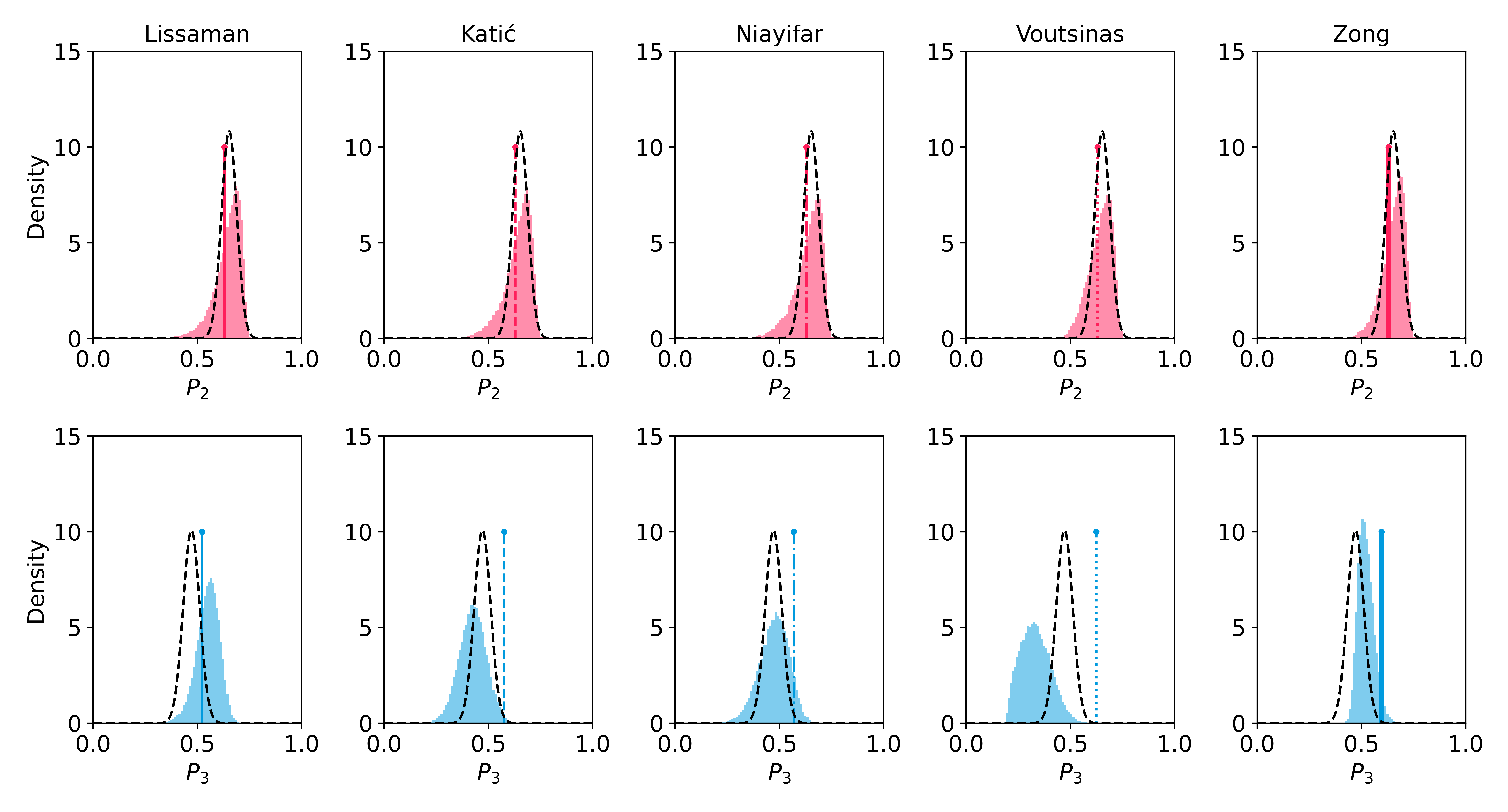}
    \caption{Power prediction for the waked turbines in the yaw scenario. The black dashed lines are the observed power production distributions, the shaded histograms are the data-driven predictions, and the shaded lines are the empirical-tuning predictions. The relative error in power prediction for the data driven method is on average about 2\% and 13\% for the means of $P_2$ and $P_3$ respectively, while the error in the predictions for the empirical tuning method is about 4\% and 22\%, respectively, with only a deterministic estimate of turbine power.
    \label{fig:prediction}
    } 
\end{figure*}

We sample from the posteriors in Fig.~\ref{fig:posteriors} to predict power production in the yaw case with $\gamma_1 = 20^\circ$, as seen in Fig.~\ref{fig:prediction} for each superposition method.
The relative predictive error of each method compared to the mean of the observation is summarized in Tab.~\ref{tab:predictions}.
For WT2, the data-driven predictions are slightly more accurate: the data-driven predictions accurately predict the mean of the observation within about 2\% for all superposition methods, while the empirical tuning predictions underpredict the mean observed power by about 3.5\%.
For WT3, the data-driven predictions are again more accurate, with an average error of about 13.5\% compared with an overprediction of 22.4\% for the empirical tuning approach.

\begin{table}
\caption{\label{tab:predictions} Relative predictive error of the parameter estimates compared with the mean observed power. `DD' refers to the data-driven method predictions and `ET' refers to the empirical tuning method predictions; the index `2' and `3' refer to the power prediction for WT2 and WT3, respectively.}
\begin{tabular}{||c|c|c|c|c||}
\hline
\textbf{Superposition} & \textbf{DD-2} & \textbf{ET-2} & \textbf{DD-3} & \textbf{ET-3} \\
\hline \hline
Lissaman & -0.9\% & -3.5\% & 16.7\% & 10.7\% \\
\hline
Kati\'{c} & -1.7\% & -3.5\% & -10.0\% & 22.2\% \\
\hline
Niayifar & -2.0\% & -3.5\% & 0.6\% & 20.6\% \\
\hline
Voutsinas & -1.4\% & -3.5\% & -29.4\% & 32.3\% \\
\hline
Zong & 0.1\% & -3.5\% & 10.7\% & 26.4\% \\
\hline
\textbf{Mean Error} & \textbf{1.2\%} & \textbf{3.5\%} & \textbf{13.5\%} & \textbf{22.4\%} \\
\hline
\end{tabular}
\end{table}

Only for the prediction of $P_3$ with the Lissaman superposition method is the empirical tuning prediction more accurate than the data-driven one, reducing the overprediction from 16.7\% to 10.7\%.
The data-driven predictions with the Niayifar method have the highest accuracy, with total error in the power prediction of the farm (i.e. $P_2$ and $P_3$ combined) less than 3\%.
It is unsurprising that the Voutsinas method results in the largest error in power prediction, considering the unusual trend in the wake expansion rates from Figs.~\ref{fig:superposition} and \ref{fig:posteriors}.
The standard deviation of the predictions for all five superposition methods match less closely than the means, with differences up to 80\% compared with the observed standard deviation.

The heightened error of the empirical tuning predictions is a combination of several factors. 
The wake-added turbulence model (Eq.~\ref{eq:prediction}), sum-of-squares turbulence superposition scheme (Eq.~\ref{eq:TI_sos}), and $k-I$ correlation (Eq.~\ref{eq:k_ti}) are all potential sources of error that do not factor into the data-driven estimation.
One could argue that while both parameter estimation approaches did not see the yaw test dataset, the data-driven calibration was still performed on the same wind farm (in a different operating scenario) and therefore its improved calibration is consistent with a reduced degree of extrapolation from the training dataset.
However, it is plausible that the data-driven method is learning a better calibration of the wake model parameters than is permitted by the model form of the empirical approach.
Further work in wake-added turbulence modeling and superposition would help to elucidate the roots of this error and potentially improve the predictive accuracy of the empirical tuning approach.

One reason for the discrepancies between the predictions and observations is that the wake models neglect some physics of turbines under yaw misalignment.
The wake model in Section \ref{sec:models} does not include any curled wake dynamics \cite{howland_wake_2016, martinez2021curled, bastankhah2022vortex}, secondary steering \cite{howland_influence_2020, king_control-oriented_2021}, or yaw-added wake recovery \cite{king_control-oriented_2021}.
Including these additional models in our wake modeling framework would alter and potentially increase the accuracy of the model predictions presented.

From these out-of-sample prediction results, we see improved predictive accuracy from the data-driven parameter estimates compared to the empirical tuning estimates. 
Averaged across all five superposition methods, the data-driven predictions reduce error compared with the empirical tuning predictions by roughly 65\% for $P_2$ and 40\% for $P_3$.
Furthermore, the data-driven method includes estimates of uncertainty in power production that the empirical tuning method is unable to produce.
The inclusion of uncertainty in $k$, which propagates to uncertainty in power, provides estimates for expected turbine performance and its variations.

\section{Conclusions}\label{sec:conc}

In this paper, we have developed a data-driven method based on Markov chain Monte Carlo to estimate the wake expansion rate of a Gaussian wake model. We use this framework to explore two key questions: (1) the effect of the wake superposition method on the parameter estimates and its contribution to model error; (2) the predictive capabilities of this data-driven approach compared to the conventional method for estimating the wake expansion rate with tuned empirical correlations.

We found that there were significant quantitative differences in the computed posterior distributions of the wake expansion rate when using different superposition methods.
Since the wake expansion rate is a parameter of the physical system and should not depend on the modeling framework, these discrepancies are a form of model error.
We emphasize that the wake superposition method should be consistent between the parameter estimation and power prediction processes and we highlight the need for more studies on wake superposition methods that are derived from first principles (e.g. Zong \cite{zong_momentum-conserving_2020} and the cumulative wake model \cite{bastankhah_analytical_2021}).

As for predictive accuracy, we found that the data-driven method, compared to the empirical tuning method, reduced error on average in a sample out of the scope of the calibration data.
In particular, the data-driven prediction based on the Niayifar superposition method resulted in less than 3\% error in predicting the mean power production of the wind farm.
The data-driven approach also provides quantification of uncertainty, which is useful for predictions in an inherently uncertain operating environment.
In combination, these two benefits make our data-driven parameter estimation framework an attractive approach to calibrate and quantify uncertainty for an engineering wake model at a site of interest.

In our experiments, we only considered conditions with a single wind speed, wind direction, and ambient turbulence intensity.
Our approach could be extended to a range of wind conditions: for each discrete combination of wind speed, wind direction, turbulence intensity, stability, and other relevant meteorological variables, the wake expansion rate could be estimated for each turbine.
While this approach would be more computationally-intensive and require a large calibration dataset (i.e. historical operational SCADA data), it could provide more accurate and robust wind farm performance estimates under a wide range of expected wind conditions compared with using ``off-the-shelf'' estimates of the wake expansion rate calibrated in limited studies.
This set of parameter estimates could inform a yaw angle optimization under uncertainty study and guide an optimal wake steering control strategy at a given wind farm site.
Furthermore, the calibration could be updated with additional measurements over time to improve performance in an open-loop framework, or it could leverage  periodic flow and power production measurements in a closed-loop control setting.

\section*{Acknowledgements}
We thank Professor Sanjiva Lele for his support and guidance throughout the development of this project.

\section*{Author Declarations}
The authors have no conflicts to disclose.

\section*{Data Availability Statement}
The data that support the findings of this study are available from the corresponding author upon reasonable request

% If in two-column mode, this environment will change to single-column format so that long equations can be displayed. 
%\begin{widetext}
%$$\mbox{put long equation here}$$
%\end{widetext}

% Fig.s should be put into the text as floats. 
% Use the graphics or graphicx packages (distributed with LaTeX2e).
% Put the label that you will use with \ref{} command in the braces of the \label{} command.
%
% \begin{figure}
% \end{figure}

% \begin{table}
% \end{tabular}
% \end{table}

%\begin{acknowledgments}
% Put your acknowledgments here.
%\end{acknowledgments}

\bibliography{references}

\section*{Appendix: Time-Average Length}

\begin{figure*}
    \centering
    \includegraphics[width=0.99\textwidth]{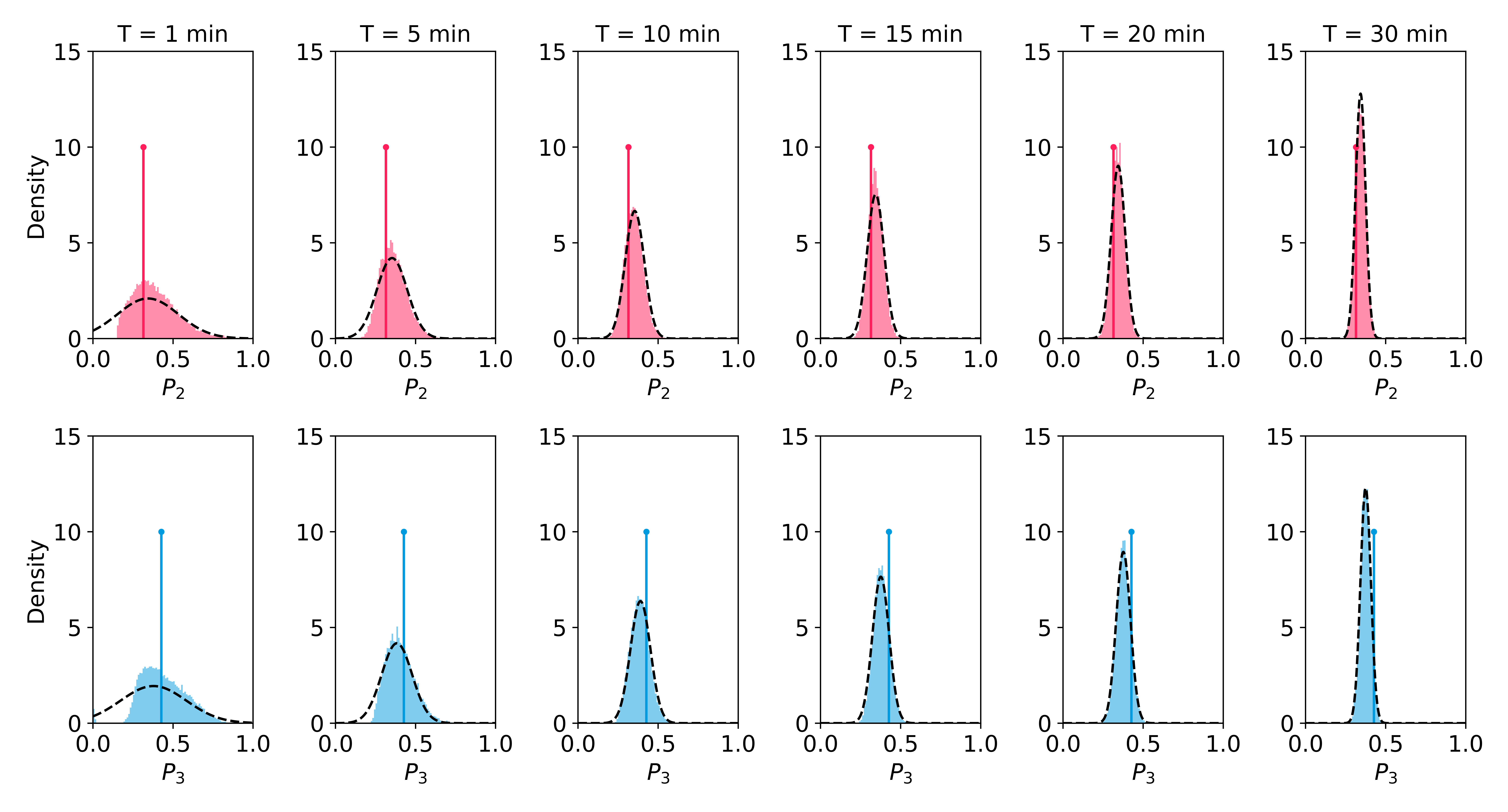}
    \caption{Power predictions of the data-driven parameter estimates (calibrated and predicted with the Zong superposition method) with different time-average lengths for the training dataset. The original dataset uses 1-minute time-averages and 10-minute time-averages were chosen for the experiments in Section \ref{sec:exp2}.
    \label{fig:time_average_training}
    } 
\end{figure*}

The raw power production dataset from the utility-scale wind farm is available in 1-minute time-averages, sampled from each turbine's onboard SCADA system.
The perfect-model power predictions are shown in Fig.~\ref{fig:time_average_training} for a range of averaging lengths from 1 to 30 minutes (approximated using ensemble averaging over independent 1-min averaged samples).
As expected, as the time-averaging length increases, the observed variability due to turbulence decreases, and the error between the predicted and observed distributions decreases.
The longer time-averages are more likely to contain all of the relevant energetic scales to resolve mean power production, by ensemble averaging over independent instantiations with approximately fixed wind conditions.
As mentioned before, the steady-state Gaussian wake model is derived assuming an infinite time average, and so the validity of our comparison to data improves when the time-averaging length is sufficiently long.

The wake model formulation contains a lower limit for the estimated power production because of two factors: (1) the physical lower bound on the wake expansion rate (Eq.~\ref{eq:limit}) and (2) the potential core length, within which there is no wake deficit prediction (Eq.~\ref{eq:core}).
Therefore, there is a lower bound for the power predictions from the steady-state Gaussian wake model.
We observe this firm boundary in $P_2$ for the 1-minute time-average length in Fig.~\ref{fig:time_average_training}.
With our assumption of a Normal distribution for the observed power, there is a non-zero probability of extremely low power production (e.g. $P_i <  20\%$) which the steady-state Gaussian wake model is unable to capture.

We select the 10-minute averaging length, which we believe results in satisfactory agreement of the data-driven power predictions in Fig.~\ref{fig:time_average_training} without constraining the variability of the observations too drastically.
At time-averaging length $T = $ 1 min, the mean error in the predictions of both the data-driven and empirical tuning methods are on the order of 10\%.
At $T = $ 10 min, the errors of the data-driven method are reduced to the order of 1\%; the empirical tuning predictions are unchanged because the mean of the observations is not changing and no calibration was done.

\begin{figure*}
    \centering
    \includegraphics[width=0.99\textwidth]{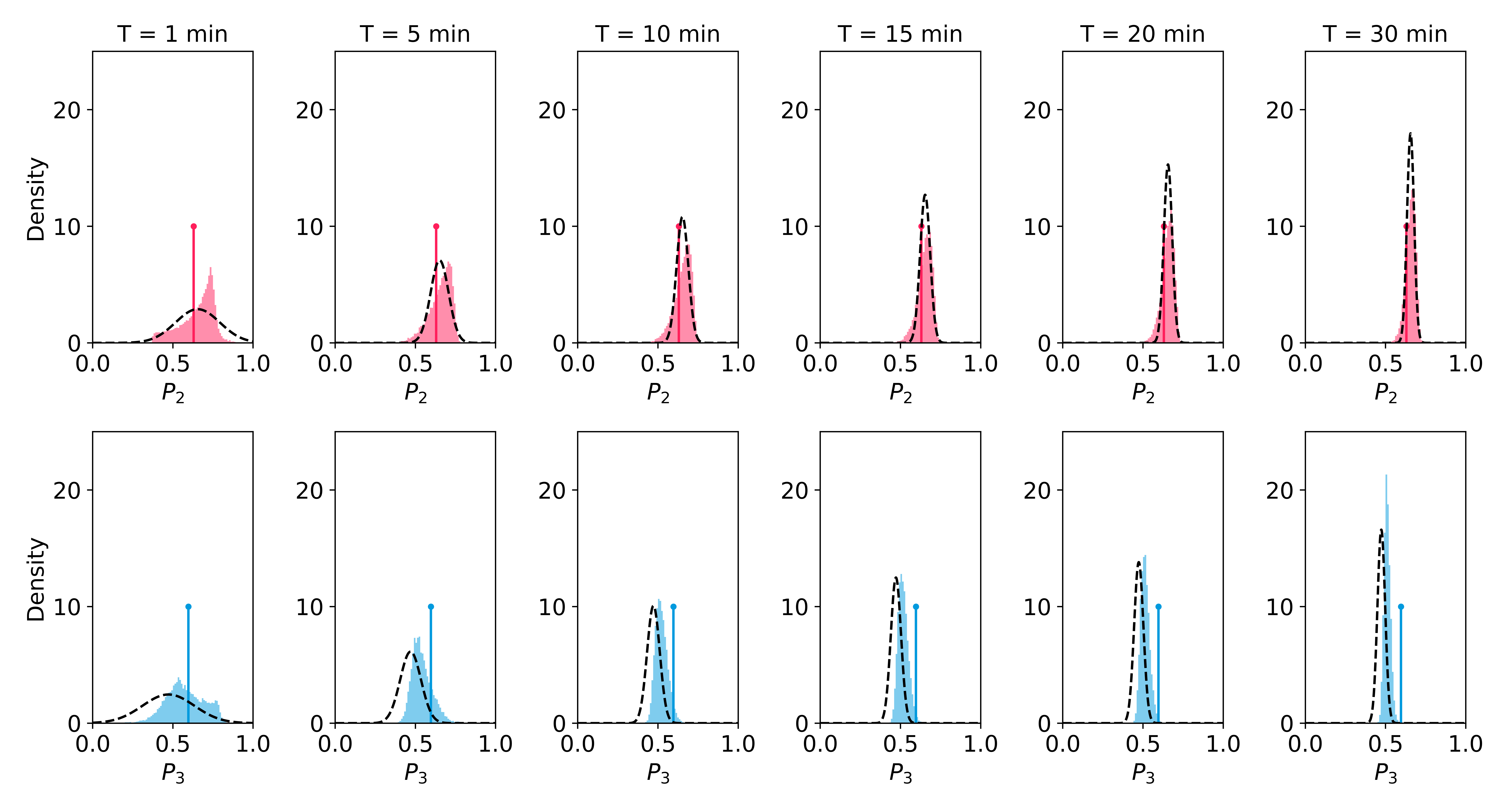}
    \caption{Power predictions in the yaw scenario compared to the test dataset with different time-average lengths (with Zong superposition used for both calibration and prediction).
    \label{fig:time_average_test}
    } 
\end{figure*}

Fig.~\ref{fig:time_average_test} illustrates the power predictions in the yaw scenario, compared to the test dataset, for the same range of time-averaging lengths.
The empirical tuning predictions have about 3\% error for $P_2$ and 25\% error for $P_3$, while the average errors for the data-driven predictions remain on the order of 0.1\% for $P_2$ and 9\% for $P_3$ for all time-average lengths $T \geq $ 5 min.
These results are included to demonstrate that our qualitative conclusions in Section ~\ref{sec:results2} are insensitive to this choice of 10-minute averages.

\end{document}